\documentclass{elsart}         

\usepackage{graphicx}
\usepackage[numbers, sort&compress ]{natbib}

\begin{document}
\begin{frontmatter}

\title{Single Particle Fragmentation in Ultrasound Assisted Impact Comminution}
\author[wittel]{Falk K. Wittel}
\corauth[cor]{Corresponding author. Tel.:+41-44-633-2871}
 \ead{fwittel@ethz.ch}
\address[wittel]{Computational Physics for Engineering Materials, Institute for Building Materials (IFB), ETH Zurich, Schafmattstrasse 6, HIF; CH-8093 Zurich}
\begin{keyword}
mpact comminution, fragmentation, DEM, ultra sonic
\end{keyword}

\begin{abstract}
Impact fragmentation is the underlying principle of comminution milling of dry, bulk solids. Unfortunately the outcome of the fragmentation process is more or less determined by the dimensionality of the impactor and its impact velocity. Since fragmentation is dominated by interfering shock waves, manipulating traveling shock waves and adding energy to the system during its fragmentation could be a promising approach to manipulate fragment mass distributions and energy input. In a former study we explored mechanisms in impact fragmentation of spheres, using a three-dimensional Discrete Element Model (DEM)\cite{carmona-wittel-etal-2008}. This work is focused on studying how single spheres fragment when impacted on a planar vibrating target.
\end{abstract} 
\end{frontmatter}
\section{Introduction}\label{sec:intro}
Fragmentation is the fundamental underlying process in many industrial comminution applications. Size reduction to desired fragment mass distributions are wanted, minimizing the energy input and process times. Single particle comminution is one of the most efficient size reduction methods, since the enormous energy losses in other processes, such as ball milling, originate from frictional inter-particle collisions \cite{tavares}. Therefore the focus of studies is on understanding and optimizing single particle comminution, most of the time by considering circular and spherical impactors. Experiments range from single or double impact of large concrete \cite{tomas-etal-99,schubert-etal-2005} or plaster \cite{arbiter-etal-69,wu-chau-yu-2004} balls over ceramic \cite{salman-etal-2002, antonyuk-etal-2006, andrews-kim-98} or glass \cite{arbiter-etal-69,andrews-kim-99,salman-gorham-2000,cheong-etal-2004} spheres in the millimeter range. Spheres of photo elastic active polymers like PMMA allowed a partial insight into the stress field dynamics during the fragmentation process \cite{majzoub-chaudhri-2000,schoenert-2004} and the crack formation inside the impactor, straightening the fact, that the problem can by described correctly only by fully tree dimensional (3D) models. Due to the violent dynamic nature of fragmentation, including multiple contacts, mostly molecular dynamics or discrete element methods (DEM) were used. However 3D simulations are rare and mainly 2D simulations, that can only describe disc fragmentation, were performed \cite{kun-herrmann-96,kun-herrmann-99-2,kun-etal-1999,khanal-etal-2004,behera-kun-etal-2005,schubert-etal-2005,thornton-etal-96}. Potapov and Campbell \cite{potapov-96,potapov-campbell-2001} introduced a 3D fragmentation simulation of spheres, composed of polyhedral particles, resulting in small system sizes, that allowed only for a rough estimate of the experimentally observed fragmentation mechanism. By using spherical particles, Thornton et.al. \cite{thornton-etal-2000,mishra-thornton-2001} could increase the particle number up to 5000, however the cohesive interactions remained quite simple. In a previous work we studied in detail the dynamics of fragmentation mechanisms during single particle impact of a system composed of an agglomeration of approximately 22000 spherical particles, interconnected by 3D beam elements \cite{carmona-wittel-etal-2008} and demonstrated the agreement with experiments.

When it comes to the technological realization via impact comminution milling, attempts to manipulate the outcome of the fragmentation process focus on process parameter like impact velocity, impact angle \cite{behera-kun-etal-2005}, target stiffness and shape \cite{schubert-etal-2005}. Fragmentation of disordered, brittle materials however proved to be a quite universal phenomenon that is mainly concerned with the impact velocity and way shock waves propagate inside the system. Experiments \cite{turcotte,oddershede-etal-1993}, as well as simulations \cite{inaoka-takayasu-96,astroem-holian-timonen-2000,astroem-etal-04,linna-astroem-2005,astroem-2006} repeatedly showed, that the outcome of the fragmentation in terms of the fragment mass and size distribution, follows a power law in the range of small fragments with exponents, that are universal with respect to the specific material or the way energy is imparted in the system.

An explanation for the universality in fragmentation and its dimensional dependency is the similarity of propagating shock waves for various materials and impactor geometries, leading to identical fragmentation mechanisms. A way to manipulate the outcome of fragmentation could be the manipulation of the shock wave configurations. The available energy for the formation of new surfaces in single particle impact is fixed by the kinetic energy of the impactor. This paper proposes the modification of shock front configurations. In medicine, destruction of kidney stones via extracorporeal shock wave lithotripsyhas has become a standard medical procedure. However for excitation, one either needs a good acoustic coupling, e.g. via fluids or a lot of time. Both is not possible in impact fragmentation, however one can excite the impactor directly at the impact, by vibrating the target. This ultrasonic assisted fragmentation is simulated using our previous model \cite{carmona-wittel-etal-2008} only with a larger number of particles and a target that is vibrating with adjustable amplitudes and frequencies. First the utilized model is recalled, before internal stress fields, energetics, damage evolution, fragmentation mechanisms, and the final outcome at various settings are compared.
\section{Model and system construction}\label{sec:model}
Since the DEM was proposed by Cundall and Struck in 1979 \cite{Cundall}, the approach had a strong attraction for the simulation of rock mechanics and brittle failure in particular. The reasons are obvious, when thinking of brittle, heterogeneous and disordered materials, that are full of defects by nature. When the ultimate strength is reached, the solid fails through the propagation of cracks, whose speed is controlled by the available energy, its flux to the failure zones and instabilities at the small scales. By representing the material via a discontinuous particle agglomeration and solving the dynamic linear and non-linear interaction of all particles, one obtains a system with complex behavior on the model scale and in particular in the crack process zones. From this, many macroscopically observed fracture phenomena naturally emerge, like size effects or crack tip instabilities with resulting crack branching and merging in dynamic propagation. Today DEM is defined as a collection of numerical methods that allow for finite displacements and rotations of discrete bodies including complete detachment \cite{Bicanic}. It is basically an explicit solution of a many body system with neighborhood search and special interaction potentials from arbitrary contact and rheological cohesive elements. Being a dynamic simulation scheme with bottom up description of the material with inherent disorder, cracking properties and crack-crack interactions naturally emerge. 
\begin{figure}[htb]
  \centering{ \includegraphics[scale=0.65]{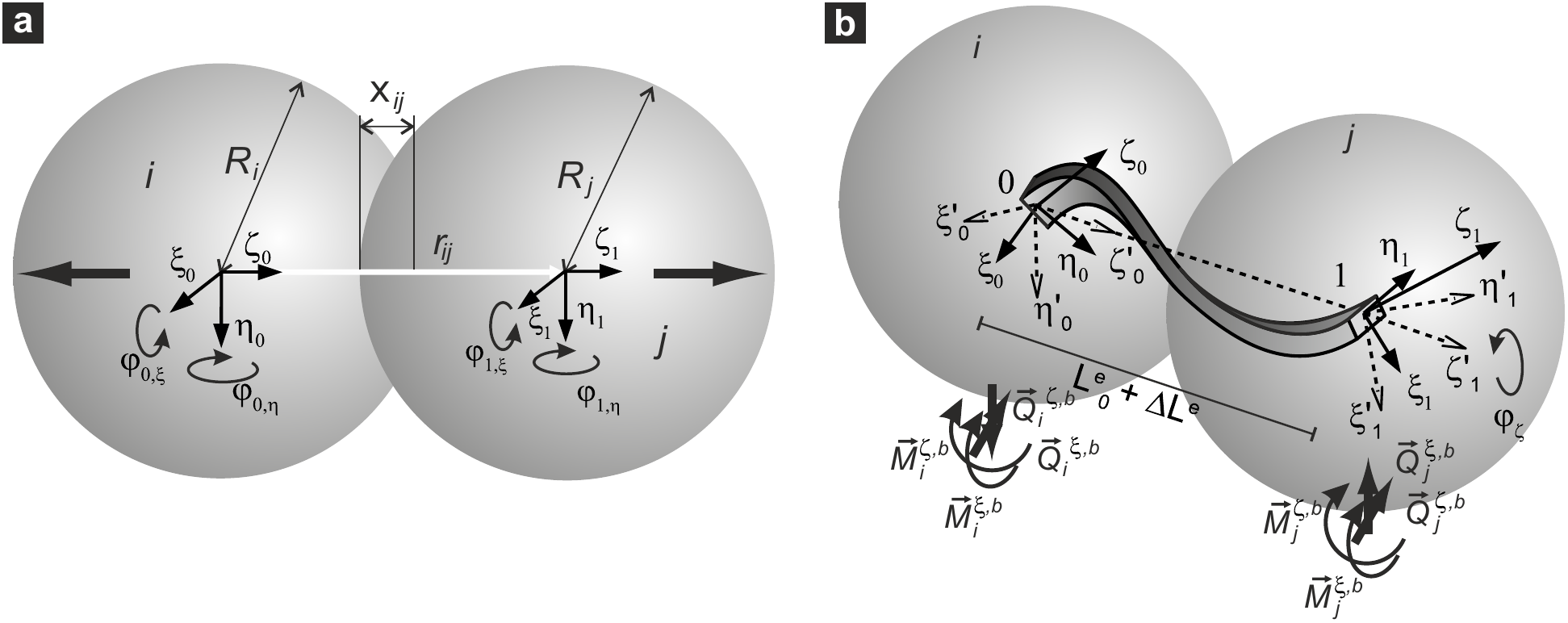} }
  \caption{\label{fig1} Inter-sphere interaction (a) and beam deformation with Euler-Bernoulli element (b).}
\end{figure}

A 3D implementation is employed, where the solid is represented by an agglomeration of bi-disperse rigid spheres. Cohesion is considered by connecting neighboring sphere centers by 3D beam-truss elements that can elongate, shear, bend and torque. The resulting force for accelerating spheres is composed of inter-sphere (see Fig.~\ref{fig1}(a)) or sphere-plane contact forces with Hertzian contact, axial forces from the truss, bending forces and moments transmitted by intact beams (see Fig.~\ref{fig1}(b)) and volumetric forces. Beam elements are allowed to fail to explicitly model damage and fracture of the solid. The utilized failure criterion on the element level considers failure due to a combination of straining and bending by comparing actual states to threshold values originating from a Weibull distribution. The material disorder is therefore considered by the  physical disorder in element breaking thresholds and topological disorder of elements. While the first one determines how the system reacts on a crack tip, namely the failure behavior, the second one is essential for obtaining realistic crack morphologies and isotropic wave propagation. Additionally damping, friction forces and torque of cohesive elements are implemented. A detailed description of the model, its calibration and verification can be found in Ref.~\cite{carmona-wittel-etal-2008}.

The system construction is a crucial step in fragmentation simulations to avoid artifacts arising from the discretization, that are difficult to detect. Namely anisotropic properties, nonuniform wave propagation or preferred crack orientations due to particle clustering or larger zones with crystalline particle arrangement have to be avoided. By using particles of slightly different sizes with diameter $d_2=0.95d_1$ of equal portions, crystalline zones can be avoided. The generation of a random agglomeration starts with an initial configuration of particles placed on a regular cubic lattice and assign random initial velocities to the particles that can move and collide in a simulation box with periodic boundaries. After this randomization step, a small central gravitational field in the center of the simulation box is activated to build one big, nearly spherical cluster of randomized spheres. After the kinetic energy has been dissipated by damping, the set of vertices is triangulated and beam elements are assigned to all edges using a Delaunay triangulation \cite{qhull}. By calculating the angular correlation of neighboring elements we could verify that crystallization is not significant, and radial alignment is not detectable \cite{carmona-wittel-etal-2008}. When the connectivity of the future beam network is found, the gravitational field is slowly removed, while the elastic beam properties are simultaneously increased. The resulting expansion reduces the contact forces. By reinitializing the beam lengths and orientations, residual stresses can be removed. The system construction is completed by trimming it to the desired shape by element removal.

Finally the system is placed at a small distance from a frictionless target plate with a 20 times higher stiffness than the spheres. The vibration of the target is considered by periodically displacing the target plate with wave length $w_t$ and amplitude $a_t$. The time evolution of the system is followed using a 6$^{th}$ order Gear predictor-corrector scheme with quaternion angle representations. Since the time increment $\Delta t$ is around 3~ns, high frequencies can be resolved quite well. For a comparison, the contact time between impactor and target is around 30~$\mu$s.
\section{Energetics and Stress Distribution}\label{sec:energetics}
\begin{figure}[htb]
  \centering{  \includegraphics[scale=0.65]{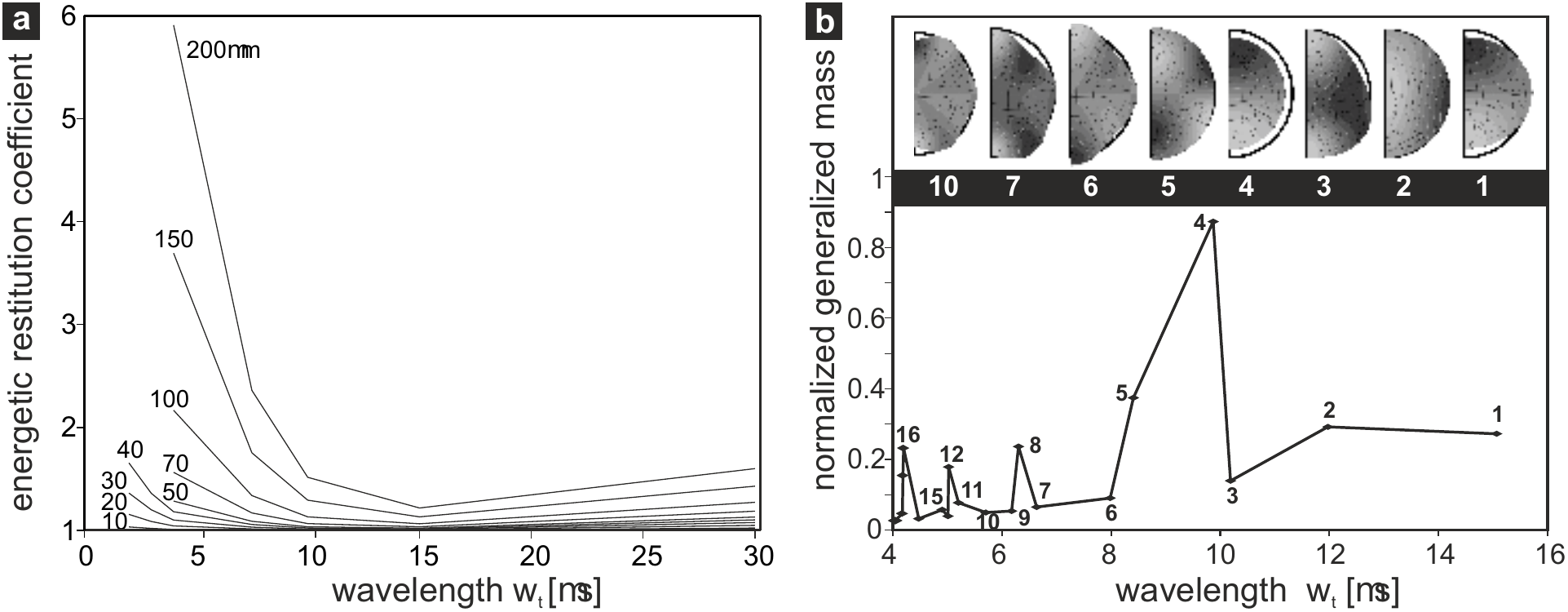} }
  \caption{\label{fig2} Energy increase due to contact stimulation of eigenmodes for amplitudes ranging from 10 to 200$\mu$m, impact velocity $v_i=145$m/s and various wave lengths (a).  In (b) the generalized mass, normalized by the total system mass is given for the first eigenmodes (1-7,10). Note that the generalized mass is defined by the product $\theta^N_\alpha M^{NM}  \theta^M_\alpha $ of the model's mass matrix $M^{NM}$  with the $\alpha-th$ eigenvector $\theta^N_\alpha$ eigenvector of the model. $N$ and $M$ refer to the degrees of freedom of the model.}
\end{figure}

When monitoring the total energy of the system during impact, energy dissipation due to damage formation, friction and damping is observed. For stationary targets the total energy will always be less or equal the initial kinetic energy. However for vibrating targets, energy is transmitted to the impactor if the wave length $w$ gets smaller than the contact time $\Delta t_c$ (see Fig.~\ref{fig2}(a)). Note that the contact time for the system at an impact speed of 145~m/s is about 31~$\mu s$ and the time for wave transmission of compression waves from the bottom to the top is about 7~$\mu$s. Therefore the number of wave packages that can be transmitted to the system is limited by the contact time.

To obtain an insight into the stress fields just before the system disintegration and the ideal increase in energy, an explicit Finite Element (FE) analysis using ABAQUS is employed. The FE model consists of quadratic axisymmetric 8-node elements, who's assigned properties originate from measurements on the DEM sample. Contact times, shock wave velocity and elastic energy show excellent agreement with the DEM simulation \cite{carmona-wittel-etal-2008}. Note that the impactor diameter is 16mm, Young's modulus is $E$=7.4GPa, density is $\rho=$1920kg/m$^3$, Poissonian number $\nu$=0.2, resulting in a longitudinal wave speed of $\approx 2200\pm 100$m/s, parameters that are in the range of lean concrete mixtures. Contact is defined as frictionless in tangential direction and hard in normal direction.

The energy of the sphere before and after impact can be compared using the FEM model to obtain a rough estimate of the energy transfer due to the vibration when failure is not present (see Fig.~\ref{fig2}(a)).
To maximize energy transfer, simulating eigenmodes is most promising. Looking at the first eigenmodes, it is evident, that only a selection can be excited by displacing the contact zone. However by exaltation with $w_t \approx 7~\mu$m, already a large number of modes is stimulated simultaneously (see Fig.~\ref{fig2}).

\begin{figure}[htb]
  \centering{  \includegraphics[scale=0.65]{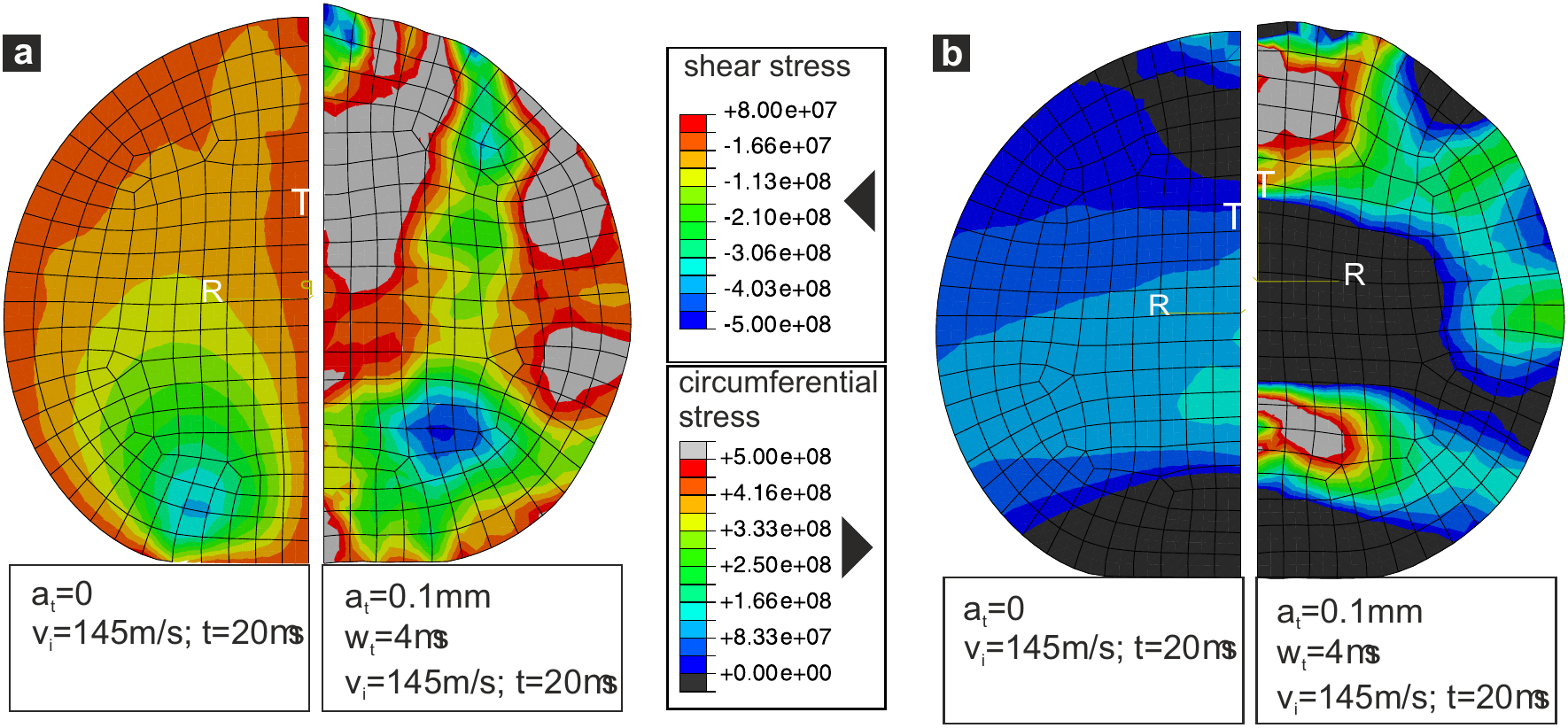}}
  \caption{\label{fig3} Shear and circumferential stresses for impact of a sphere at 145 m/s with static and vibrating targets. (Color online)}
\end{figure}

The FEM simulations also help to clarify the stress fields and regions with stored elastic energy since any breakage is determined by the stress field. The stress fields for single particle impact with statics targets are quasi static in the sense, that reflected shock waves \cite{schoenert-2004} do not dominate the overall stress field. During the contact, stress magnitudes simply rise and fall \cite{khanal-etal-2004,andrews-kim-98,carmona-wittel-etal-2008,kienzler-schmitt-90}. Maximal shear stresses are found close to the change from the curvature to the flattened region at the contact. Inside the sample a biaxial stress state is found with tensile circumferential stresses and strong longitudinal compression (comp. Fig.~\ref{fig3}(a)).

With vibrating targets, constantly new elastic waves emerge from the contact and propagate through the sphere. The stress fields therefore are strongly modified, leading necessarily to diverse fragmentation mechanism. For small wave lengths, the contact is separated and closed repeatedly (comp. Fig.~\ref{fig3}(b)).
\section{Fragmentation}
In impact comminution, the degree of fragmentation and its evolution strongly depends on the impact velocity. To emphasize the effect of the US-assisted fragmentation, the focus is put on low velocity impact with a velocity just above the characteristic fragmentation threshold. First the effect of the vibrating target on characteristic fragment sizes is analyzed, before the fragmentation mechanisms are studied in detail, that are responsible for the functional shape of the fragment mass distribution, analyzed thereafter.
\subsection{Fragment sizes}
\begin{figure}[htb]
  \centering{  \includegraphics[scale=0.65]{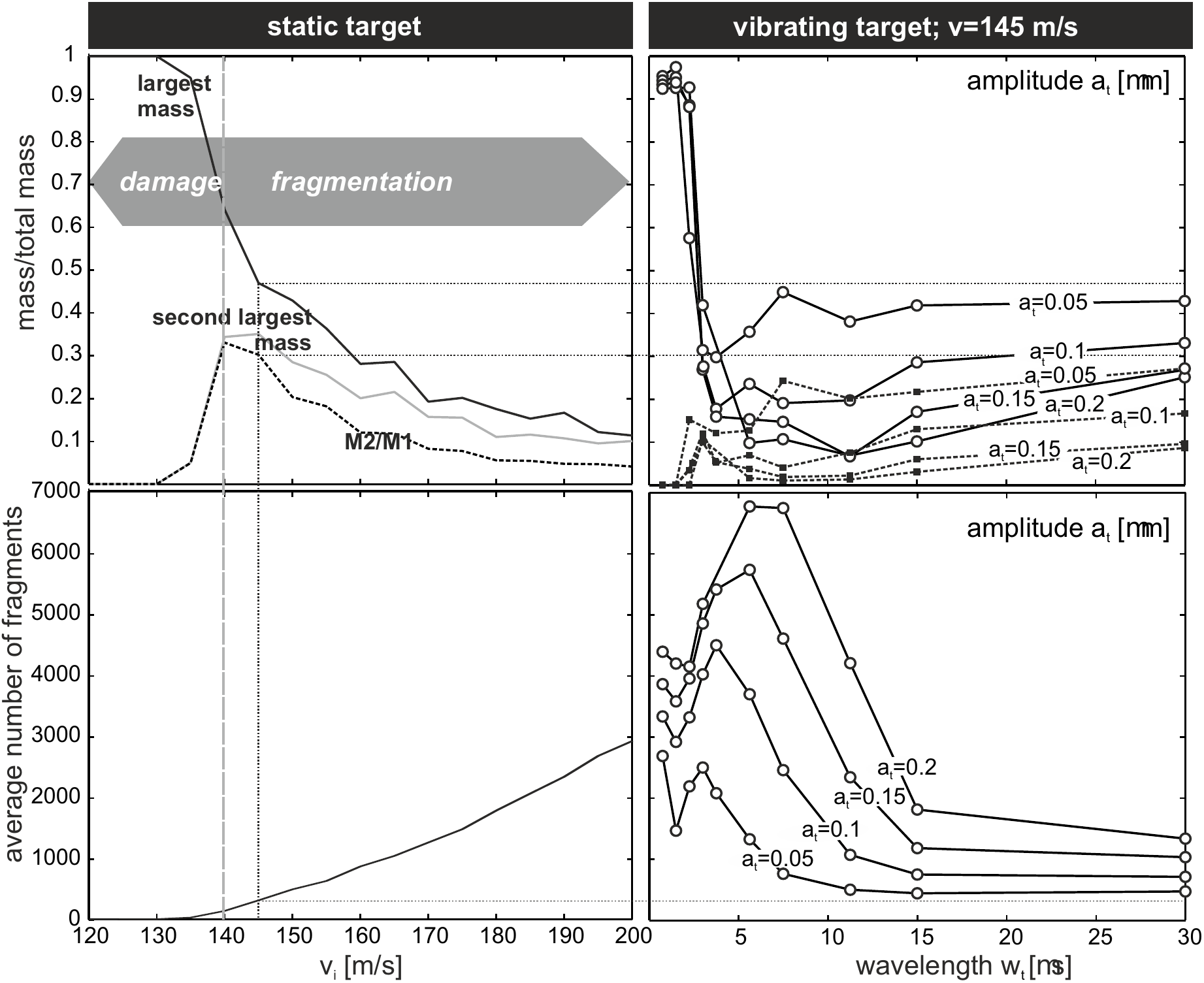}}
  \caption{\label{fig4} Scaling for largest, 2$^{nd}$-largest and average fragment mass (top row) and number of fragments (bottom row) as function of impact velocity for stationary targets (left column) and for US-assisted impact at $v_i$=145m/s with diverse wave length $w$ and amplitude $a$ (right column).}
\end{figure}

In an ideal milling process, fragment sizes would be uni-disperse. However in reality a fragmentation leads to a wide distribution of fragment sizes, that can span several orders of magnitude. If the velocity of an impactor is gradually increased, first micro damage at the contact zone is observed, however the integrity of the impactor is remained. Note that already in this damage regime, small fragments consisting of up to a few elementary particles are released. As soon as the fragmentation threshold is reached, the largest fragments break into smaller ones, leading to an equalization of the largest with the 2$^{nd}$ largest clusters as shown in Fig.~\ref{fig4}. As a representation value for the average fragment size the quotient $M_2/M_1$ of moments $M_k=\sum_i^{N_f} M_i^k-M_{max}^k$ is used, omitting the largest cluster \cite{kun-herrmann-96}.

Fragment sizes can be compared for increasing frequency starting from one wave length per contact time (30$\mu$s) up to 40. Fig.~\ref{fig4} demonstrates that the maximum and average fragment size reduction due to US is significant. As a matter of fact maximum fragment sizes with $a=$0.15mm and $w_t=$12$\mu$s at an impact velocity of 145 m/s correspond to an impact velocity of more than 200m/s for static targets and the average size to an even higher one. The fragment numbers (Fig.~\ref{fig4} bottom) exhibit a similar drastic increase. Interestingly ultrasound can not only promote, but also prohibit fragmentation (see Fig.\ref{fig4}(c) $w_t<$3$\mu$s). By only looking at the fragment numbers for high frequencies, the opposite would have been expected, however most of the energy is dissipated by grinding up the contact zone and frictional particle interactions and is no longer available for crack propagation. This examplefies that a closer look on occurring fragmentation mechanisms is necessary. 
\subsection{Fragmentation Process and Mechanisms}
\begin{figure}[htb]
  \centering{  \includegraphics[scale=0.65]{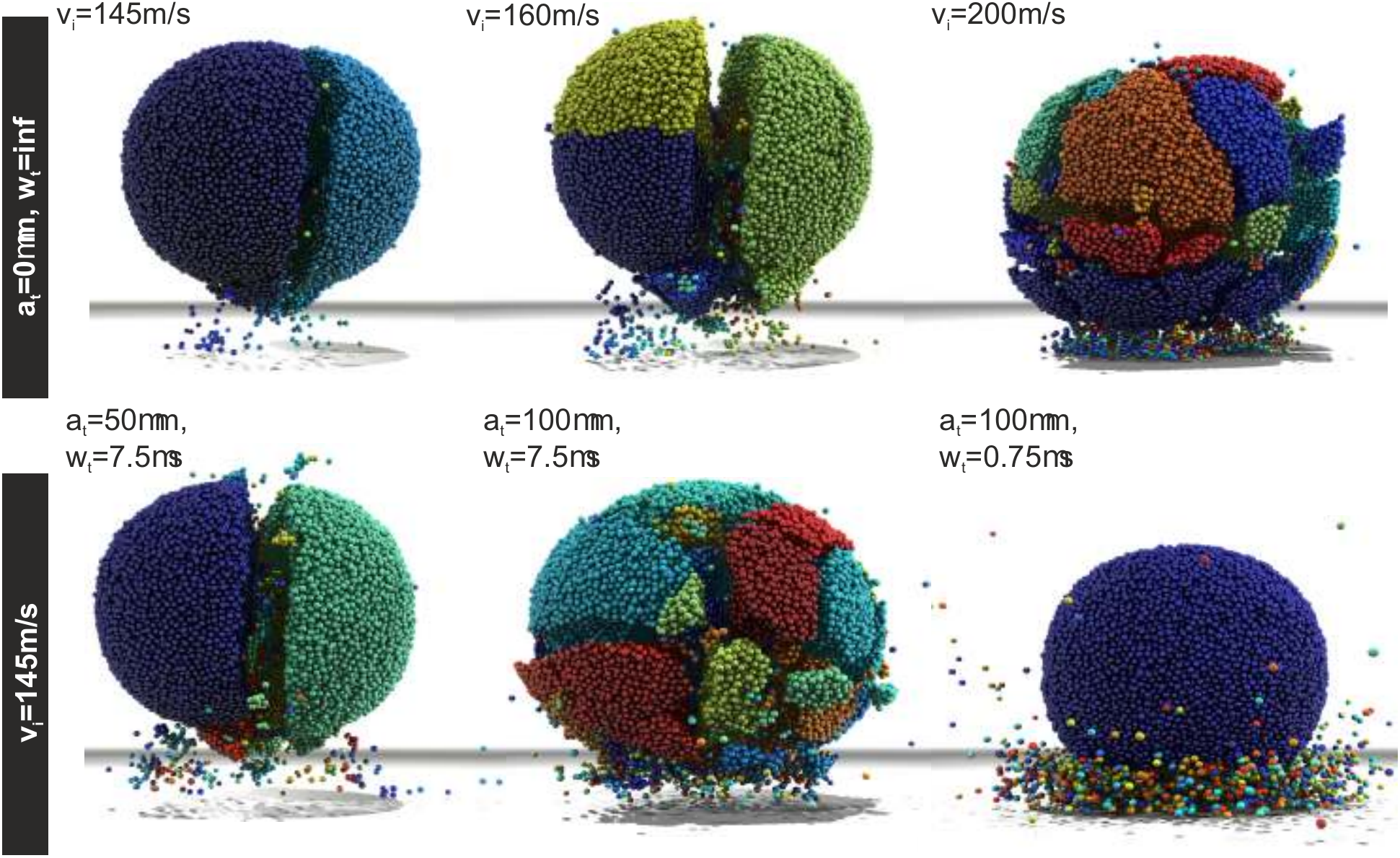}}
  \caption{\label{fig5}  Final fragmented stage for static targets. Intensities represent different clusters. (Color online)}
\end{figure}

\begin{figure}[htb]
  \centering{  \includegraphics[scale=0.65]{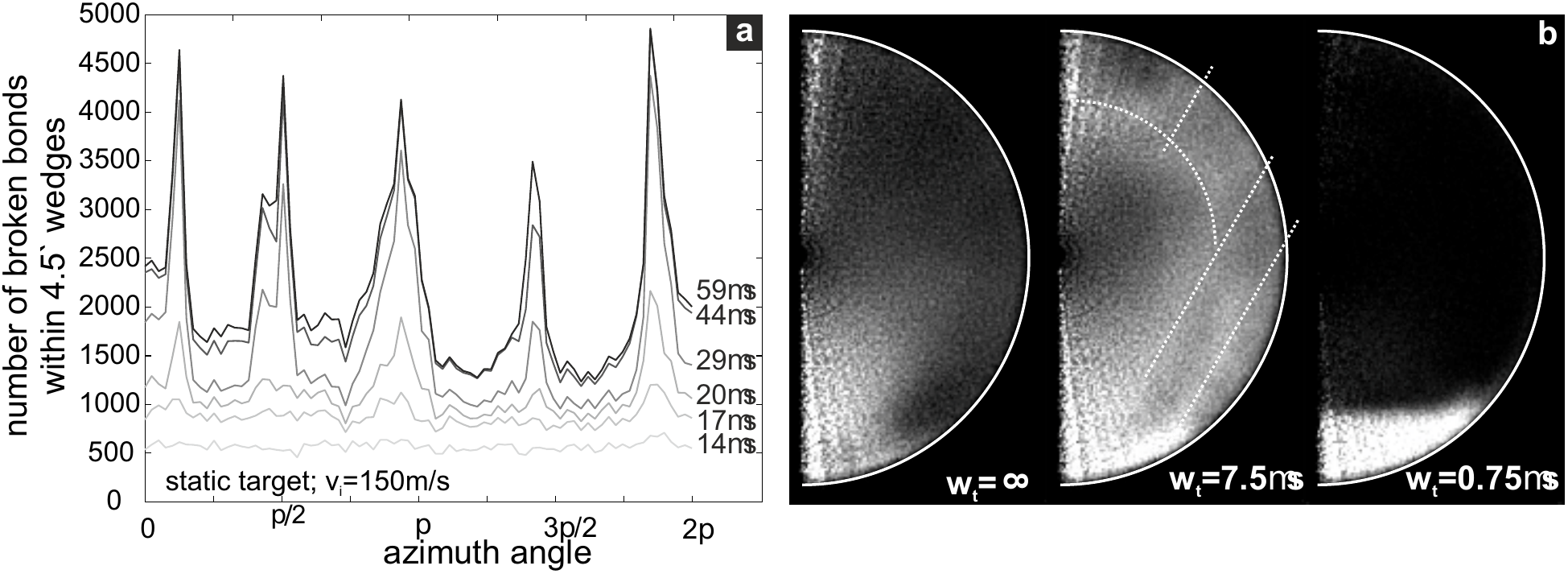}}
  \caption{\label{fig6} Evolution of the angular distribution of broken elements (a) and average damage maps for various wave length at an amplitude $a$=150$\mu$m (b).}
\end{figure}

Single particle impact against static targets was subject of experimental \cite{arbiter-etal-69,andrews-kim-98,majzoub-chaudhri-2000, salman-gorham-2000,tomas-etal-99,wu-chau-yu-2004,schoenert-2004,salman-etal-2002,antonyuk-etal-2006} and numeric \cite{carmona-wittel-etal-2008,thornton-etal-2004, kienzler-schmitt-90,thornton-etal-96,potapov-96,mishra-thornton-2001} investigations before. For low velocities uncorrelated damage initiates about D/4 from plane inside the sphere in the region with the biaxial stress state described before. This zone gets weakened by micro cracks. Around the weakened core, the material has high circumferential tensile stresses in a ring shaped zone, where meridional cracks originate. Since the number of meridional cracks depends on the stress rates, we concluded, that their stress release fields interact like in ring fragmentation \cite{mott,molinari-etal-2007}. Once initiated, meridional cracks can grow from the inside to the outside with energy dependent angular separation of wedge shaped fragments provided enough energy is imparted \cite{carmona-wittel-etal-2008}. Also a ring of broken bonds is observed, forming a cone, basically by failure due to shear at the contact zone. For higher velocities, oblique cracks further fragment the wedge shaped fragments due to diverse stress states. To summarize, for a static target quasi periodic sharp meridional cracks splitting the impactor (see Fig.~\ref{fig6}a), some fragments at the impact cone and a few one particle fragments are observed. Damage is mainly localized to form large cracks (see Figs.~\ref{fig5},\ref{fig6}). 

\begin{figure}[htb]
  \centering{  \includegraphics[scale=0.65]{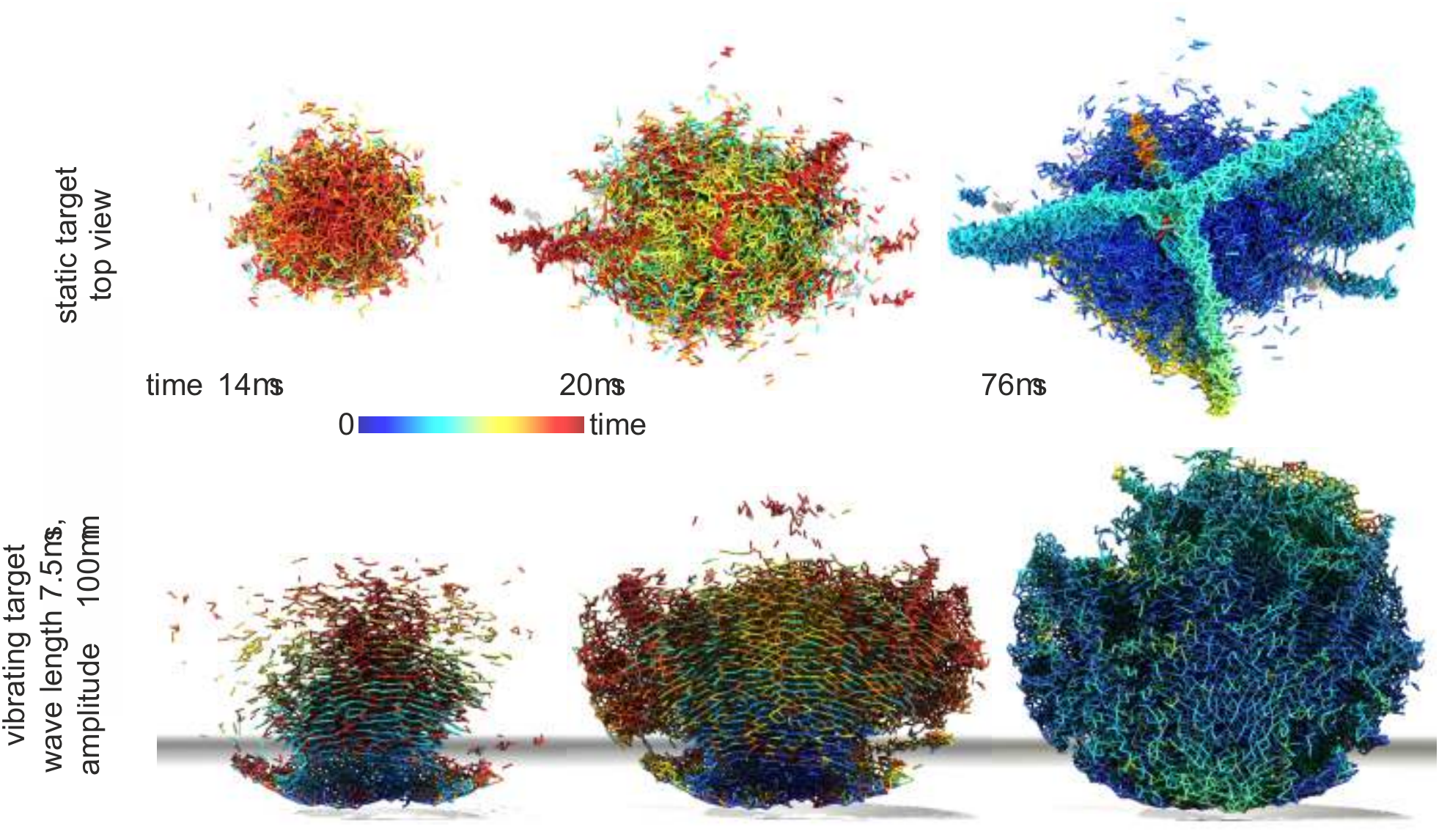}}
  \caption{\label{fig7} Comparison of damage evolution for static and vibrating targets for $v_i=145$m/s. Broken beams are colored corresponding to their failure time from black (first failure) to white (time of snapshot). Single broken beams are pruned for a better visibility of main cracks. (Color online)}
\end{figure}

With a vibrating target the emerging fragmentation mechanisms change, depending on amplitude and wave length (see Fig.\ref{fig7}). Already for small amplitudes of 50$\mu$m the impact cone gets further fragmented, new fragments form and damage zones widen. Opposite to the case of the static target at identical impact velocity, also inside fragments damage is dispersedly distributed, simplifying further fragmentation e.g. in a secondary comminution step. In Fig.\ref{fig5} a case with higher amplitude of 100$\mu$m is compared with impact against a static target at high velocity. It is visible by the naked eye, that fragment shapes and consequently fragmentation mechanisms differ. For static targets, secondary fragmentation of wedge shaped fragments dominates. The case of US-assisted fragmentation is characterized by one front of fractures that grow simultaneously from the bottom to the top, leaving the fragmented system behind (see Fig.~\ref{fig7}).

It is interesting to note, that for high frequencies of vibrating targets, fragmentation of the bulk impactor is prohibited by a protecting layer of shattered material that forms in the contact region by abrasion or surface erosion and acts as a highly dissipative granulate. The frequencies for this rather sharp transition e.g. in the maximum fragment size (see Fig.~\ref{fig4}) mark a limit for ultra sound assisted comminution.
\subsection{Mass distributions of fragments}
\begin{figure}[htb]
  \centering{  \includegraphics[scale=0.65]{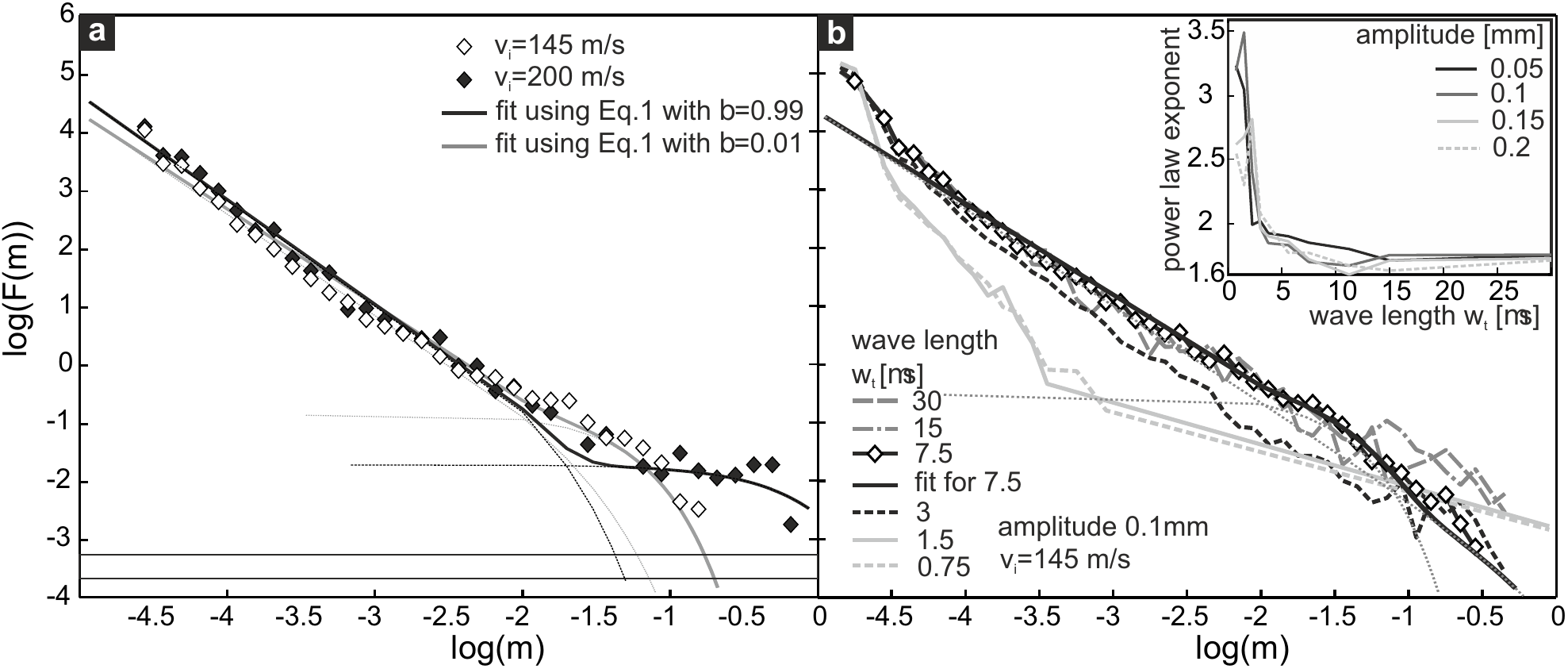}}
  \caption{\label{fig8} Fragment mass distributions for static targets (a) and vibrating ones (b). The inset shows the dependences of the power law exponent $\beta$ for small fragment masses with the transition to the shattered phase.}
\end{figure}

To sample the outcome of a fragmentation process, size or mass distributions are among the most important characteristic quantities. The most frequently used form for expressing fragment size distributions are the cummulative mass of fragments with at least one enclosing diameter smaller than the size of the wire line used in experimental sieving measurements. Unfortunately large fragments are well represented this way on the expense of the small ones. To obtain also accurate results in the low fragment mass range, the fragment mass probability density function $F(m)$ has to be calculated from the cumulation of fragments formed over all realizations of identical control parameter sets, using logarithmic binning for fragment masses normalized by the total mass of the system, $m$. This corresponds to a form of size distribution as the number density of fragments with normalized mass inside a certain mass range. The functional form of fragment mass distributions (FMD) $F(m)$ was first described in power-law form by Trucotte \cite{turcotte}, with an universal exponent. Different exponents were found for fragmentation of objects of lower dimensionality like shells \cite{falk2,falk3,falk4}, plates or rods. Attempts to explain the outcome of the fragmentation process by statistical geometric processes however all lead to FMD of some kind of exponential and not power-law form. \AA str\"om et.al. \cite{astroem-etal-04} showed, that the dynamics of the fragmentation process has to be considered even in minimal models to explain realistic FMD. They proposed a relation that is composed of two parts, the first one being a dynamic branching-merging process known for dynamic crack propagation, and a second one originating from the Poissonian nucleation process of the first dominating cracks, namely
 \begin{equation}\label{eq:btd1}
   F(m) \sim  (1-\beta) m^{-\tau} \exp\left(-m/\bar{m}_0\right) + \beta \exp\left(-m/\bar{m}_1\right).
 \end{equation}
The $\beta$ parameter controls the relative importance of the branching-merging and Poissonian nucleation process, while the exponent $\tau$ only depends on the dimensionality of the system, which is in this case $\tau_{3D}=(2D-1)/D=5/3$. $\bar{m}_0$ and $\bar{m}_1$ are cut of values of the respective parts. For static targets a good fit is obtained for $v_i=145$m/s when $\beta=0.99$ and for $v_i=200$m/s when $\beta=0.01$ (see Fig.~\ref{fig8}). This is in agreement with the observation, that just above the fragmentation threshold the Poissonian fracture nucleation is relevant, while for high velocities dynamic branching-merging mechanisms dominate.
To answer the question, which mechanisms dominate in US-assisted fragmentation, the fragment mass distributions are evaluated for various wave lengths $w_t$. For $w_t\approx 30\mu$s, the distributions for static targets hold quite well, while with increasing wave length, $\beta$ has to be reduced to around 0.01 to obtain a good fit. Therefore also in US-assisted fragmentation, branching-merging mechanisms start to dominate. When a threshold wave length of $\approx 3\mu$s is reached, the exponent of the power law part jumps from 1.67 to significantly higher values, since the contact region is in the shattered phase.
\section{Conclusions}
It was shown in a realistic 3D DEM simulation, that ultra sonic assisted impact comminution has a huge potential. The energy transfer into the impactor can be realized during the short contact time with vertically vibrating targets. Not only the energy increase, but also resulting multiple shock fronts lead to higher degree of comminution and fragmentation mechanisms, that differ from those of classical impact comminution. The fragment number that can be obtained with already quite small amplitudes of 50$\mu$m is comparable to an increase of impact energy via velocity of about 80\%. If the frequencies are above the first eigenmodes, and wave lengths are smaller than the overall contact time, a considerable amount of energy is pumped into the impactor and available for crack propagation. Multiple shock fronts lead to a more uniform distriubiton of energy and a higher degree of comminution. This goes along with a change in observable fragmentation mechanisms. Impact fragmentation against static targets lead to oblique cracks and secondary fragmentation of wedge formed framents. Impact against vibrating targets lead to a strong fragmentation of the cone and contact zone. Fragents form  by local crack branching-merging along a fragmentation front from the bottom to the top. If freqencies get to high, the system is pushed back into the damage regime due to a protective layer of shattered material that forms at the contact zone. The required frequencies and amplitudes for the chosen mechanical parameters, that are in the range of lean concrete mixtures, are within the range of "off-shelf" US-actuators and transducers. To estimate the specific energetic gain of US-assisted fragmentation however, one needs to consider the whole chain, including the efficiency of transducers. 

\end{document}